\documentclass[prl,twocolumn,amsmath,superscriptaddress,amssymb]{revtex4-2}

\usepackage{graphicx}
\usepackage{dcolumn}
\usepackage{bm}
\usepackage{amssymb}
\usepackage{amsmath}
\usepackage{wasysym}
\usepackage{color}
\usepackage{times}

\usepackage{placeins}
\usepackage{epstopdf}

\newcommand{\ket}[1]{\left|\,#1\,\right\rangle}

\usepackage[normalem]{ulem}

\makeatletter
\renewcommand*\env@matrix[1][c]{\hskip -\arraycolsep
  \let\@ifnextchar\new@ifnextchar
  \array{*\c@MaxMatrixCols #1}}
\makeatother
\begin{document}

\title{
Tomographic Imaging of Orbital Vortex Lines in Three-Dimensional Momentum Space
}

\author{T. Figgemeier}
\thanks{These authors contributed equally to the present work.}
\affiliation{Experimentelle Physik VII and W\"urzburg-Dresden Cluster of Excellence ct.qmat, Universit\"at W\"urzburg, Am Hubland, D-97074 W\"urzburg, Germany}

\author{M. \"Unzelmann}
\thanks{These authors contributed equally to the present work.}
\email{muenzelmann@physik.uni-wuerzburg.de}
\affiliation{Experimentelle Physik VII and W\"urzburg-Dresden Cluster of Excellence ct.qmat, Universit\"at W\"urzburg, Am Hubland, D-97074 W\"urzburg, Germany}

\author{P. Eck}
\thanks{These authors contributed equally to the present work.}
\affiliation{ITPA and W\"urzburg-Dresden Cluster of Excellence ct.qmat, Universit\"at W\"urzburg, Am Hubland, D-97074 W\"urzburg, Germany}

\author{J. Schusser}
\affiliation{Experimentelle Physik VII and W\"urzburg-Dresden Cluster of Excellence ct.qmat, Universit\"at W\"urzburg, Am Hubland, D-97074 W\"urzburg, Germany}

\author{L. Crippa}
\affiliation{ITPA and W\"urzburg-Dresden Cluster of Excellence ct.qmat, Universit\"at W\"urzburg, Am Hubland, D-97074 W\"urzburg, Germany}

\author{J. N. Neu}
\affiliation{Department of Chemical and Biomedical Engineering, FAMU-FSU College of Engineering
	Tallahassee, FL 32310, USA}
\affiliation{National High Magnetic Field Laboratory, Tallahassee, FL 32310, USA}

\author{B. Geldiyev}
\affiliation{Experimentelle Physik VII and W\"urzburg-Dresden Cluster of Excellence ct.qmat, Universit\"at W\"urzburg, Am Hubland, D-97074 W\"urzburg, Germany}

\author{P. Kagerer}
\affiliation{Experimentelle Physik VII and W\"urzburg-Dresden Cluster of Excellence ct.qmat, Universit\"at W\"urzburg, Am Hubland, D-97074 W\"urzburg, Germany}

\author{J. Buck}
\affiliation{
	Institut f\"ur Experimentelle und Angewandte Physik, Christian-Albrechts-Universit\"at zu Kiel, D-24098 Kiel, Germany}
\affiliation{Ruprecht Haensel Laboratory, Kiel University and DESY, D-24098 Kiel and D-22607 Hamburg, Germany}

\author{M. Kall\"ane}
\affiliation{
	Institut f\"ur Experimentelle und Angewandte Physik, Christian-Albrechts-Universit\"at zu Kiel, D-24098 Kiel, Germany}
\affiliation{Ruprecht Haensel Laboratory, Kiel University and DESY, D-24098 Kiel and D-22607 Hamburg, Germany}

\author{M. Hoesch}
\affiliation{
	Deutsches Elektronen-Synchrotron DESY, D-22607 Hamburg, Germany}

\author{K. Rossnagel}
\affiliation{	Institut f\"ur Experimentelle und Angewandte Physik, Christian-Albrechts-Universit\"at zu Kiel, D-24098 Kiel, Germany}
\affiliation{Ruprecht Haensel Laboratory, Kiel University and DESY, D-24098 Kiel and D-22607 Hamburg, Germany}
\affiliation{	Deutsches Elektronen-Synchrotron DESY, D-22607 Hamburg, Germany}

\author{T. Siegrist}
\affiliation{Department of Chemical and Biomedical Engineering, FAMU-FSU College of Engineering
	Tallahassee, FL 32310, USA}
\affiliation{National High Magnetic Field Laboratory, Tallahassee, FL 32310, USA}

\author{L.-K. Lim}
\affiliation{Zhejiang Institute of Modern Physics, Department of Physics, Zhejiang University, Hangzhou, Zhejiang 310027, People’s Republic of China}

\author{R. Moessner}
\affiliation{Max Planck Institute for the Physics of Complex Systems and W\"urzburg-Dresden Cluster of Excellence ct.qmat, Noethnitzer Strasse 38, D-01187 Dresden, Germany}

\author{G. Sangiovanni}
\affiliation{ITPA  and W\"urzburg-Dresden Cluster of Excellence ct.qmat, Universit\"at W\"urzburg, Am Hubland, D-97074 W\"urzburg, Germany}

\author{D. Di Sante}
\affiliation{Department of Physics and Astronomy, Univerity of Bologna, I-40127 Bologna, Italy}

\author{F. Reinert}
\affiliation{Experimentelle Physik VII and W\"urzburg-Dresden Cluster of Excellence ct.qmat, Universit\"at W\"urzburg, Am Hubland, D-97074 W\"urzburg, Germany}

\author{H. Bentmann}
\affiliation{Experimentelle Physik VII and W\"urzburg-Dresden Cluster of Excellence ct.qmat, Universit\"at W\"urzburg, Am Hubland, D-97074 W\"urzburg, Germany}
\affiliation{Center for Quantum Spintronics, Department of Physics, NTNU Norwegian University of Science and Technology,
NO-7491 Trondheim, Norway}

\date{\today}

\begin{abstract}


We report the experimental discovery of orbital vortex lines in the three-dimensional (3D) band structure of a topological semimetal. Combining linear and circular dichroism in soft x-ray angle-resolved photoemission (SX-ARPES) with first-principles theory, we image the winding of atomic orbital angular momentum, thereby revealing — and determining the location of — lines of vorticity in full 3D momentum space. Our observation of momentum-space vortex lines with quantized winding number establishes an analogue to real-space quantum vortices, for instance, in type-II superconductors and certain non-collinear magnets. These results establish multimodal dichroism in SX-ARPES as an approach to trace 3D orbital textures. Our present findings particularly constitute the first imaging of non-trivial quantum-phase winding at line nodes and may pave the way to new orbitronic phenomena in quantum materials.

\end{abstract}
\maketitle

The geometric structure of Bloch wave functions in a periodic lattice may host non-trivial textures in momentum space underpinning the unique properties of topological quantum matter \cite{novoselov2005two,Hasan:10.11,Qi:11,Burkov:11,heikkila2015,Fang:15,Schnyder:16,bradlyn2016,fang2016topological,Haldane:17,Wen:17,Armitage:18,RevModPhys.91.015006,rhim2020,Wilde2021}. While pointlike topological defects, the celebrated Weyl points, have been extensively studied \cite{Lv:15,Xu:15,Uenzelmann:2021}, higher dimensional structures such as vortex lines have proven to be harder to pin down.
However, given the importance of line vortices in real space --- particularly prominent in superfluids \cite{volovik2003}, superconductors \cite{AbrikosovVortex},
and indeed the early universe \cite{KibbleString,vilenkin1994} 
--- it may seem surprising that such structures in reciprocal space have thus far entirely eluded detection. This is  despite the fact that their possible existence is theoretically firmly established:       
pseudospin vortex rings have recently been predicted to emerge in 3D topological semimetals \cite{volovik2003,PhysRevLett.118.016401}, and they may be associated with topological quantum transport properties \cite{Kim2018,Hirose2020, Kim2022}.
More broadly, exhibiting phenomena such as (Berry) flux quantization and quantum phase winding, momentum space vortex lines are in some respects analogous to magnetic vortices in type-II superconductors and therefore a central paradigm in condensed matter physics.
%
%
%
\begin{figure*}[t]
\includegraphics[width=\textwidth]{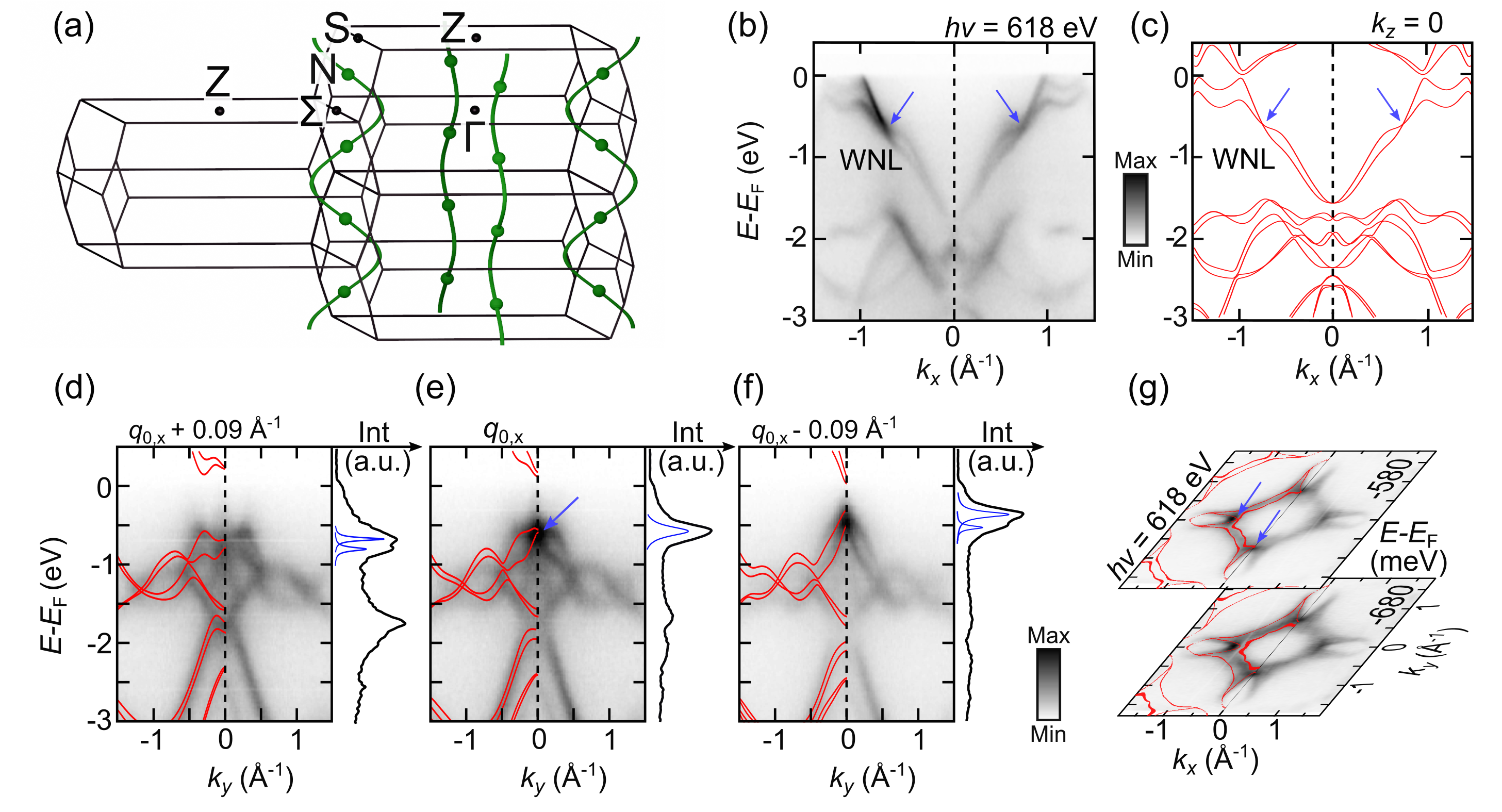}
\caption{Almost movable Weyl nodal line in TaAs. {(a) Sketch of the TaAs Brillouin zone with a symmetry-enforced almost movable Weyl nodal line (WNL), denoted in green, as predicted in Ref.~\cite{Hirschmann2021}. The WNL interconnects adjacent N points (green dots). (b,c) ARPES data and DFT calculation along the $\Gamma$-$\Sigma$ direction. A band crossing corresponding to the WNL is highlighted with blue arrows. (d,e,f) ARPES data and DFT calculations along $k_\text{y}$ taken at selected $k_x$ near the location of the WNL. The energy distribution curves (EDC) at $k_\text{y}=0\, \text{\AA}$ supports a vanishing band splitting at the WNL. (g) ARPES constant-energy-cuts recorded in the $\Gamma$-$\Sigma$ plane ($h\nu=618\,\text{eV}$) at energies near the WNL (blue arrows). 
}
}
\label{fig1}
\end{figure*}
%
%
%
%

The fact that their observation has so far remained elusive
is presumably in part due to the experimental challenge of imaging vorticity across extended volumes of 3D $k$-space, requiring a complete 3D  map of the wave function properties.
This we have now achieved with the necessary phase-sensitive and momentum-resolved contrast via linear and circular dichroism in soft X-ray angle-resolved photoelectron spectroscopy (SX-ARPES) \cite{jung2020black,Beaulieu:20,Uenzelmann:2021,Yen2023,Brinkman2024}.
Our central result is an image of 3D reciprocal space that unambiguously exhibits topological vortex lines in the orbital texture of TaAs. In recent years, this compound has attracted broad attention, especially for hosting Weyl points \cite{huang:15,Weng:15,Xu:15,Lv:15,yang:15,Uenzelmann:2021}. 
However, the topological features exposed in this work are entirely distinct from the previously discussed (Weyl point) physics. In particular, we have now detected line vortices of orbital angular momentum (OAM) that exhibit a new class of Weyl nodal lines \cite{Hirschmann2021} pinned to the vortex cores.
%

In the following, we first analyze the band dispersion around a two-fold spin-degenerate band crossing using SX-ARPES and density functional theory (DFT) bulk band structure calculations (see Supplementary Material (SM) for details on the experimental and theoretical methods).
The focus will then shift to the properties of the wave functions of the corresponding eigenstates. More precisely, we will investigate the 3D momentum dependence of the atomic OAM using dichroic SX-ARPES supported by state-of-the-art simulations of the photoemission intensity, DFT, and an effective minimal model. 
Finally, we utilize dichroism under systematically varied experimental parameters to achieve a full vectorial tomography of the orbital vortex texture.

Fig.~\ref{fig1}(b) displays SX-ARPES data along $\Gamma\Sigma$ obtained at $h\nu=618\,\mathrm{eV}$, demonstrating excellent agreement with the corresponding calculation in Fig.~\ref{fig1}(c). For our analysis, we will focus on the topmost valence band which extends approximately $1.5\,\mathrm{eV}$ below $E_\mathrm{F}$. This band exhibits a notable splitting resulting from spin-orbit coupling (SOC) and broken inversion symmetry, as confirmed by both our calculations and high-resolution data (see Supplementary Note 1 for further analysis). Interestingly, we observe a band crossing along the $\Gamma\Sigma$ path, denoted by blue arrows in Figs.~\ref{fig1}(b-c).
Around the observed crossing at $q_{0}$ the band dispersion is highly anisotropic, and our data and calculations clearly reveal splittings along both in-plane momentum directions (Fig.~\ref{fig1}(d-g)).     
\begin{figure*}
\includegraphics[width=\textwidth]{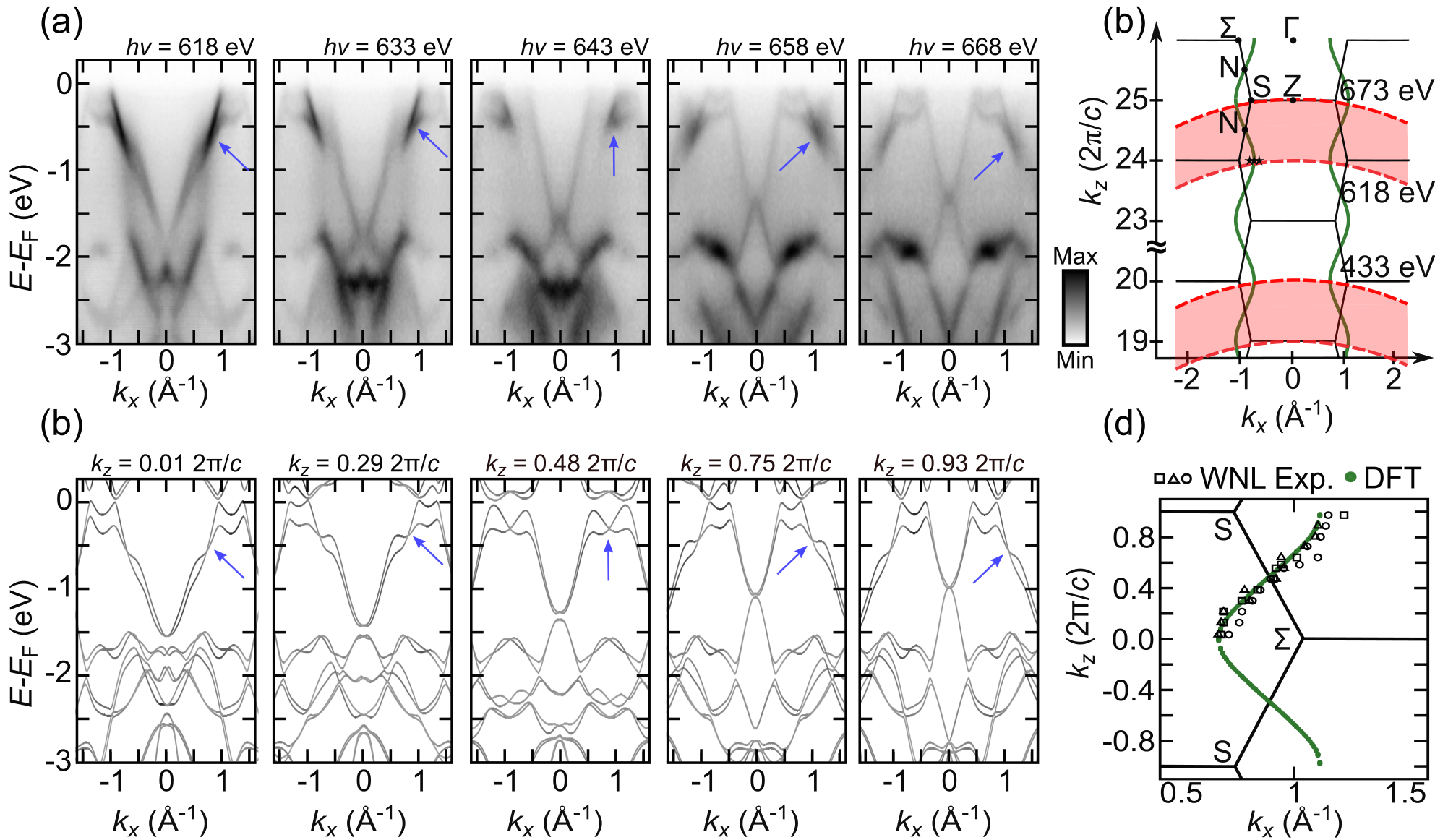}
\caption{3D evolution of the Weyl nodal line. (a) ARPES data sets along the $\Gamma$-$\Sigma$ direction for various $k_\text{z}$ throughout the Brillouin zone. Data sets are symmetrized with respect to the $\Gamma$ point. (b) DFT calculations at $k_\text{z}$-values matching the data sets in (a). Blue arrows mark the position of the WNL. (c) Bulk Brillouin zone structure of TaAs. Red dashed lines represent measurements at fixed photon energy h$\nu$. Data sets were taken in the red-shaded area. The stars represent the cuts shown in Figs.~1 d-f. (d) Position of the WNL based on DFT (green line) and h$\nu$-depdendent ARPES data (data points) in the upper red-shaded area of (c). 
}
\label{fig2}
\end{figure*}

Tracing $q_{0}$ along the out-of-plane momentum $k_z$, our DFT calculations reveal a Weyl-type --- i.e. two-fold spin-degenerate --- nodal line (WNL) that connects neighboring N-points through the $\Gamma \mathrm{ZN} \Sigma$ mirror planes, as depicted in Figs.~\ref{fig1}(a) and \ref{fig2}(b,d).
This confirms its classification as a so-called \textit{almost movable} WNL, a new class of topological feature theoretically predicted to occur in certain non-symmorphic crystals from space group $I4_1 md$ (\#109) \cite{Hirschmann2021}.
To experimentally verify this $k_z$ evolution, we conducted photon-energy-dependent measurements. We obtained data sets in two energy intervals, as illustrated in Fig.~\ref{fig2}(c). 
Representative data sets in Fig.~\ref{fig2}(a) are shown and compared to calculations at corresponding out-of-plane momenta (Fig.~\ref{fig2}(b)), achieving excellent agreement. Notably, a gradual shift of the band crossing towards larger $k_x$ is observed as $k_z$ varies from the zone center ($h\nu=618\,\mathrm{eV}$, $k_z\approx 0$) to the ZS equi-$k_z$ plane ($h\nu=668\,\mathrm{eV}$, $k_z\approx 2\pi /c$).
Thus, our measurements and calculations reveal a significant undulation of the WNL with $k_z$, where the band crossing modulates between type-II (tilted) and type-I (untilted).
Importantly, we consistently observe abrupt modulations in the photoemission intensity along $k_x$ at the band crossing for all photon energies, providing an additional spectroscopic signature of the WNL. Taken together, we plot the WNL in Fig.~\ref{fig2}(d) as extracted from the experimental data (see also Supplementary Note 1) in good agreement with the overlaid calculations. 
%
%
%
%

\begin{figure*}[t]
\includegraphics[]{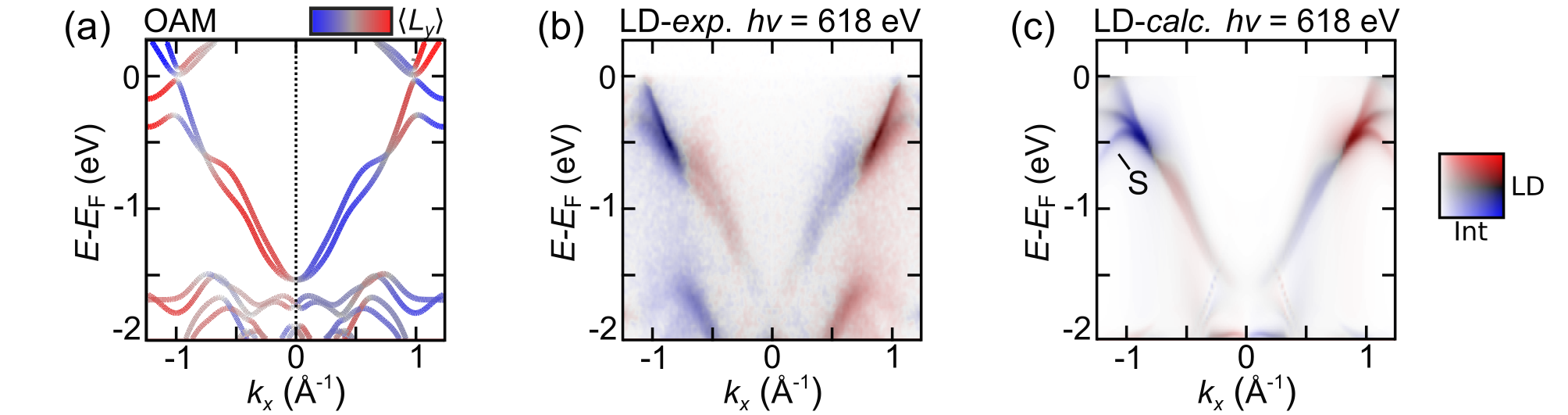}
\caption{Linear dichroism and orbital angular momentum at the WNL. (a) OAM $L_y$, as obtained from DFT.
(b) Measured linear dichroism (see text) at $h\nu = 618\,\mathrm{eV}$ corresponding to an out-of-plane momentum of $k_z\approx 0$. 
(c) Same as in (b), but obtained from a one-step photoemission calculation. Overall, the LD shows a sign change at the WNL crossing point in agreement to OAM (see SM Fig.~S6 for LD data at a different photon energy corresponding to the equivalent $k_z$). A surface resonance --- obtained only in the LD calculation in (c) --- is marked by $S$.  
}
\label{fig3}
\end{figure*}

Having established the dispersion, we will now focus on the properties of the wave functions aiming to understand the orbital texture in the vicinity of the WNL. 
To this end,
we calculated the orbital contributions to the Bloch states $\Psi_k$. Along the $k_x$ direction, the states are predominantly formed from Ta $d_{xz}$ and $d_{z^2}$~/~$d_{x^2-y^2}$ orbitals. Due to broken inversion symmetry, these orbitals hybridize with complex phases: 
$$\ket{\Psi_k}\propto  \big( \sqrt{3}\ket{d_{z^2}}  + \ket{d_{x^2-y^2}} \big) + {i\gamma(\mathbf{k})}\cdot \ket{d_{xz}}. 
$$ 
As a result, the states $\Psi_k$ carry a finite atomic orbital angular momentum (OAM) $\langle L_y \rangle \propto \gamma(k)$. Importantly, $\gamma(k)$ undergoes a sign switch at the WNL (see Supplementary Note 3), leading to a reversal of OAM (Fig.~\ref{fig3}(a)).     
\\
To address this aspect experimentally,
we consider the linear dichroism (LD) in the ARPES intensity distribution (Figs.~\ref{fig3}(b)). The LD is defined as $\mathrm{LD}=I(k_x)-I(-k_x)$ when employing p-polarized or circularly polarized light incident in the $xz$ plane. Similar to circular dichroism (CD) \cite{Park:12,Park:12_2}, the LD can be associated with OAM \cite{Uenzelmann2019}. Specifically, assuming a weak $k_x$ dependence of the photoelectron final state, leading terms in the LD are proportional to $\gamma(k)$ (see Supplementary Note 2). 
\\
The measured LD shows good agreement with the calculated OAM and reverses sign across the WNL, confirming the predicted OAM reversal (Figs.~\ref{fig3}(b)). 
To further substantiate this finding, we conducted one-step photoemission calculations (Fig.~\ref{fig3}(c)). The results provide a striking match with the experimental data, confirming that the observed LD reversal at the WNL is a genuine effect of the photoemission matrix element. A minor exception is the appearance of a surface resonance in the calculation that is not observed in the experimental data. Comparison of LD distributions along $\Gamma\Sigma$ at different photon energies (see Supplementary Fig.~S6), which both belong to the same $k_z$, indicate only a modest influence of the photoemission final state. Both the calculations and experimental data consistently display a reversal of the LD at the WNL, affirming minor contributions of trivial geometry or final-state effects but rather sensitivity of LD to the phase $\gamma(k)$ and to the OAM.

\begin{figure*}[t]
\includegraphics[]{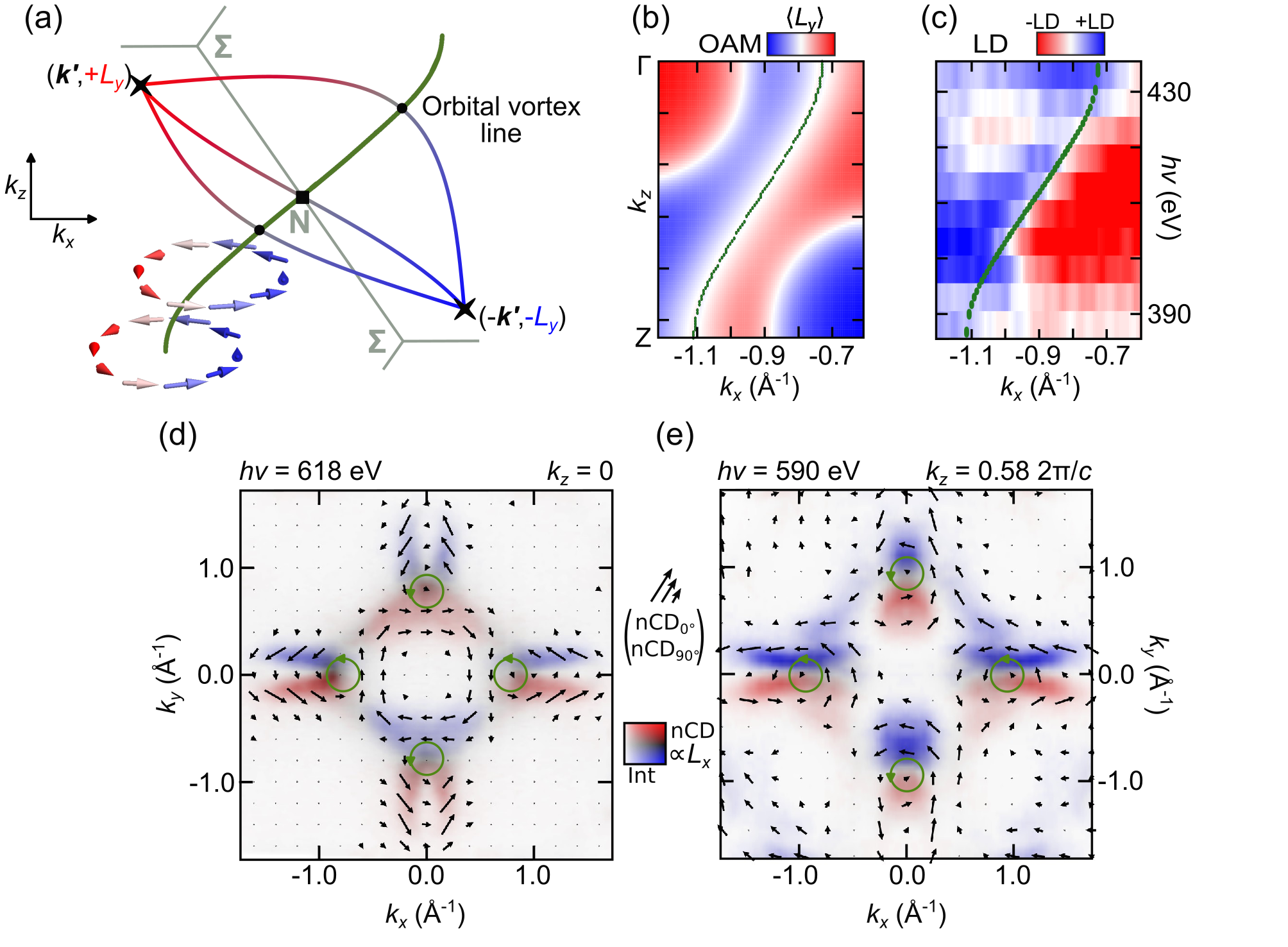}
\caption{{Orbital vortex line in three-dimensional momentum space.} (a) Illustration of symmetry-enforced formation of orbital vortex lines (OVL).  
DFT calculations of the OAM vortex configuration are depicted at different $k_\mathrm{z}$-values. (b) DFT calculation of the OAM in the $\Gamma \mathrm{ZN} \Sigma$ mirror plane.
The green line indicates the calculated position of the WNL, matching the trajectory on which the OAM vanishes. 
(c) Measured LD, integrated over the energy interval of the OVL, in dependence of photon energy $h\nu$, sampling the sign change of the LD for different $k_z$-values throughout one BZ.
(d,e) Normalized circular dichroism momentum maps obtained at photon energies $h\nu=618\,\mathrm{eV}$ and $h\nu=590\,\mathrm{eV}$ corresponding to out-of-plane momenta of $k_z\approx 0$ and $k_z\approx 0.58\,2\pi /c$, respectively. The constant energy contours are taken at the respective energy of the nodal points, i.e., $-0.58\,\mathrm{eV}$ ((d)) and $-0.38\,\mathrm{eV}$ ((e)). The dichroic pattern (red-blue) reflects OAM component $L_x$. The vectors $(\mathrm{nCD}_{0^\circ},\mathrm{nCD}_{90^\circ})$ are obtained from taking the $\mathrm{nCD}_{0^\circ}(k_x,k_y)$ and the same data set rotated by $90^\circ$. This symmetry operation ($\mathcal{R}_{90^\circ}\colon (k_x,k_y) \mapsto (-k_y,k_x)$) also transforms the in-plane OAM as $\mathcal{R}_{90^\circ}\colon L_x(k_x,k_y) \mapsto L_y(-k_y,k_x)$. 
The observed vector field around the nodal point momenta, highlighted by the green circles, resembles the OAM vortex winding.
}
\label{fig4}
\end{figure*}

From basic symmetry considerations, it is evident that the $\Gamma \mathrm{ZN} \Sigma$ mirror planes must host symmetry-enforced lines with vanishing OAM, $L = 0$ (see Fig.~\ref{fig4}(a) and Supplementary Note 3). In Fig.~\ref{fig4}(b) we present the band-resolved OAM texture within the $k_x k_z$ mirror plane. Remarkably, the OAM flips sign on a curved trajectory $\bold{q_0}(k_z)$, where $L = 0$, that perfectly aligns with the WNL (green line). Our LD data as a function of photon energy (Fig.~\ref{fig4}(c)) further validate the $k_z$-dependent undulation of the OAM sign reversal, consistent with the observed evolution of the WNL in Fig.~2. 
\\
To understand this crucial observation, we investigate the OAM texture in the vicinity of the WNL. For this purpose, we consider the Hamiltonian:
\begin{equation}
\mathcal{H}_\mathrm{OVL}(\bold{q})=\mathcal{H}_0+ H_\mathrm{ISB} (\bold{q}),
\end{equation}
where $\bold{q}= \bold{k}-\bold{q}_\mathrm{0}$ (see Supplementary Note 3). $\mathcal{H}_0$ describes the unperturbed situation and the second account for the formation of OAM $\bold{L}$ due to broken inversion symmetry. In first order, the latter can be expressed as
\begin{equation}
\mathcal{H}_\mathrm{ISB}(\bold{q}) \approx  
-\alpha_{y,x} q_{y} L_x + (\alpha_{x,y} q_x - \alpha_{z,y} q_z) L_y +\alpha_{y,z} q_y L_z ~.
    \label{H_ORE}
\end{equation}
The coefficients $\alpha_{i,j}$ are real-valued parameters and quantify the energy scale associated with the breaking of inversion symmetry. The resulting OAM texture exhibits a vortex enclosing the line $\bold{q_0}(k_z)$. Although details depend on the $k_z$-dependent parameters $\alpha_{i,j}$, the OAM vortex universally carries a winding number $\nu = 1$.
Our DFT calculations confirm these findings (see Fig.~\ref{fig4}(a) and S8) and further show that $\alpha_{y,z}$ is small for all $k_z$, implying that $\bold{L}$ has predominantly in-plane orientation.\\
There is one final ingredient missing in order to obtain the fully-fledged WNL, which is spin-orbit coupling. This lifts spin degeneracy implicit in the band structure obtained from Eq.~\ref{H_ORE}.
The non-trivial OAM winding texture, however, remains unaffected by the presence of SOC (see also Supplementary Note 3).
In essence, the line $\bold{q_0}(k_z)$ serves a precursor of the WNL, around which $H_\mathrm{SOC}\propto \bold{L}\cdot \bold{S}$ induces a band splitting $\Delta E \propto L$ that vanishes at $\bold{q_0}$.
%
%

To gain experimental access to the OAM vortex, we consider the normalized circular dichroism (nCD) in $k_x$-$k_y$ momentum maps. As shown recently, CD-ARPES enables detailed mapping of the in-plane momentum space OAM texture \cite{Uenzelmann:2021,Schusser2022}.
In Fig.~\ref{fig4}(d,e), we consider data sets at two photon energies, corresponding to out-of-plane momenta $k_z\approx 0$ and $k_z\approx 0.58\,2\pi /c$, with characteristically distinct $k_\parallel$-dispersion (compare also Fig.~\ref{fig2}).
\\
Within our experimental geometry, the nCD pattern is sensitive to the in-plane OAM component $L_x$ \cite{Uenzelmann:2021} (Supplementary Note 2). This signal --- referred to as $\mathrm{nCD}_{0^\circ}$ --- is represented by the red-blue color code in Fig.~\ref{fig4}(d,e).
In addition, four-fold rotation symmetry in the crystal structure of TaAs allows us to reconstruct the full vectorial in-plane OAM texture. This means rotation of the sample by $90^\circ$ transforms the in-plane OAM as $L_x(k_x,k_y) = L_y(-k_y,k_x)$. This in turn provides access to the orthogonal component $\mathrm{nCD}_{90^\circ} \propto L_y$ and we can determine the in-plane OAM vector fields plotted as arrows in Fig.~\ref{fig4}(d,e).
We directly observe vortex textures around the four nodal points (also Supplementary Fig.~S8), as predicted from our model Hamiltonian (Eq.~\ref{H_ORE}) and DFT calculations (Fig.~\ref{fig4}(a) and Supplementary Fig.~S8). This applies to both exemplary out-of-plane momenta considered, with the positions of the vortex cores $\bold{q_0}$ varying according to the undulation of the nodal line (Figs.~\ref{fig2}(d) and \ref{fig4}(b,c)).
\\
\\
In conclusion, our experiments provide a complete three-dimensional tomographic image of orbital vortex lines in the electronic structure of a topological quantum material. These vortices emerge as line defects in the 3D Bloch wave-function manifold, and their existence is closely tied to the band topology. Specifically, our measurement of orbital winding directly reveals a topological charge, linked to the presence of a line node.


We established multimodal dichroism in $k_z$-resolving SX-ARPES as a powerful approach to trace the full orbital pseudospin texture in 3D materials \cite{Schueler2022}.
The capabilities of dichroic photoemission tomography may pave the way for the exploration of exotic band topologies arising from line nodes, including Nexus textures and non-Abelian topological invariants \cite{heikkila2015,wu2019non}, which have so far evaded the context of realistic quantum materials.
Moreover, the extension to time-resolved experiments aiming to study the dynamics of orbital textures will be an intriguing pathway based on our present results.

Finally, we point out that our finding of orbital vortices raises the prospect of exploring novel quantum transport phenomena, which may result from non-trivial OAM textures. In particular, our results highlight the potential of investigating 3D quantum materials in the context of orbitronics \cite{Bernevig2005,Go2021, Choi2023}.

%

\section{Acknowledgments}
This work is funded by the Deutsche Forschungsgemeinschaft (DFG, German Research Foundation) through SFB 1170 'ToCoTronics', the W\"urzburg-Dresden Cluster of Excellence on Complexity and Topology in Quantum Matter --\textit{ct.qmat} Project-ID 390858490 - EXC 2147, and RE1469/13-2. The Research Council of Norway (RCN) supported H. B. through its Centres of Excellence funding scheme, project number 262633, ``QuSpin'', and RCN project number 323766. 
The research leading to these results has received funding from the European Union’s Horizon 2020 research and innovation program under the Marie Skłodowska-Curie Grant Agreement No. 897276.
We acknowledge DESY (Hamburg, Germany), a member of the Helmholtz Association HGF, for the provision of experimental facilities. Parts of this research were carried out at PETRA III and we would like to thank Kai Bagschik, Jens Viefhaus, Frank Scholz, J\"orn Seltmann and Florian Trinter for assistance in using beamline P04. Funding for the photoemission spectroscopy instrument at beamline P04 (Contracts 05KS7FK2, 05K10FK1, 05K12FK1, 05K13FK1, 05K19FK4 with Kiel University; 05KS7WW1, 05K10WW2 and 05K19WW2 with W\"urzburg University) by the Federal Ministry of Education and Research (BMBF) is gratefully acknowledged. J.S. would like to acknowledge J\'{a}n Min\'{a}r for providing a computational cluster at NTC, University of West Bohemia. We gratefully acknowledge the Gauss Centre for Supercomputing e.V. (www.gauss-centre.eu) for funding this project by providing computing time on the GCS Supercomputer SuperMUC at Leibniz Supercomputing Centre (www.lrz.de). J. N. and T. S. acknowledge support from the National Research Foundation, under Grant No. NSF DMR-1606952. The crystal synthesis and characterization was carried out at the National High Magnetic Field Laboratory, which is funded by the National Science Foundation (NSF DMR-2128556) and the State of Florida.

\section{AUTHORS CONTRIBUTIONS}
TF and  M\"U performed the experiments, with support from HB, BG, PK, JB, MH, MK, and KR. TF and M\"U analyzed the experimental data supported by PK. PE performed the first-principles calculations. JS performed the one-step photoemission calculations. JNN, TF and TS synthesized and characterized the TaAs samples. M\"U and LKL developed the model Hamiltonian with support from TF and PK. All authors contributed to the interpretation and discussion of the results. TF, M\"U and HB wrote the manuscript with contributions from PE, LC, JS, GS, DdS, LKL, RM, and FR. M\"U and HB conceived and planned the project.




%
%

\end{document}